\documentclass[12pt,preprint]{aastex}

\slugcomment{accepted by AJ}

\shorttitle{Rapid-Response Spectrophotometry of NEOs}
\shortauthors{Mommert et al.}

\begin{document}


\title{First Results from the Rapid-Response Spectrophotometric Characterization of Near-Earth Objects using UKIRT}


\author{M. Mommert\altaffilmark{1}}

\author{D. E. Trilling\altaffilmark{1}}

\author{D. Borth\altaffilmark{2}}

\author{R. Jedicke\altaffilmark{3}}

\author{N. Butler\altaffilmark{4,5}}

\author{M. Reyes-Ruiz\altaffilmark{6}}

\author{B. Pichardo\altaffilmark{7}}

\author{E. Petersen\altaffilmark{8,1}}

\author{T. Axelrod\altaffilmark{9}}

\and

\author{N. Moskovitz\altaffilmark{10}}


\email{michael.mommert@nau.edu}

\altaffiltext{1}{Department of Physics and Astronomy, Northern Arizona
  University, Flagstaff, AZ 86001, USA}
\altaffiltext{2}{Deutsches Forschungszentrum f\"{u}r K\"{u}nstliche Intelligenz (DFKI), D-67663 Kaiserslautern, Germany}
\altaffiltext{3}{Institute for Astronomy, University of Hawaii at Manoa, Honolulu, HI 96822, USA}
\altaffiltext{4}{School of Earth and Space Exploration, Arizona State University, Tempe, AZ 85287, USA}
\altaffiltext{5}{Cosmology Initiative, Arizona State University, Tempe, AZ 85287, USA}
\altaffiltext{6}{Universidad Nacional Aut\'{o}noma de M\'{e}xico, Instituto de Astronom\'{i}a, Ensenada, B.C. 22860, M\'{e}xico}
\altaffiltext{7}{Instituto de Astronom\'{i}a, Universidad Nacional Aut\'{o}noma de M\'{e}xico, Ciudad Universitaria, D.F. 04510, M\'{e}xico}
\altaffiltext{8}{Institute for Geophysics, University of Texas at Austin, Austin, TX 78758, USA}
\altaffiltext{9}{Steward Observatory, University of Arizona, Tucson, AZ 85721, USA}
\altaffiltext{10}{Lowell Observatory, Flagstaff, AZ 86001, USA}


\begin{abstract}
  Using the {\it Wide Field Camera} for the United Kingdom Infrared
  Telescope, we measure the near-infrared colors of near-Earth objects
  (NEOs) in order to put constraints on their taxonomic
  classifications. The rapid-response character of our observations
  allows us to observe NEOs when they are close to the Earth and
  bright. Here we present near-infrared color measurements of 86 NEOs,
  most of which were observed within a few days of their discovery,
  allowing us to characterize NEOs with diameters of only a few
  meters. Using machine-learning methods, we compare our measurements
  to existing asteroid spectral data and provide probabilistic
  taxonomic classifications for our targets. Our observations allow us
  to distinguish between S-complex, C/X-complex, D-type, and V-type
  asteroids. Our results suggest that the fraction of S-complex
  asteroids in the whole NEO population is lower than the fraction of
  ordinary chondrites in the meteorite fall statistics. Future data
  obtained with UKIRT will be used to investigate the significance of
  this discrepancy.
\end{abstract}





\section{Introduction}

Near-Earth Objects (NEOs) are Solar System bodies whose orbits bring
them close to the Earth's orbit. NEOs constitute a short-lived
small-body population that is replenished by different asteroid
populations, most of which lie within the asteroid main belt, and by
comets from the outskirts of the Solar System \citep[see,
e.g.,][]{Bottke2002}. Some NEOs pose a direct threat to Earth, as has
been recently seen in the Chelyabinsk airburst
\citep{Brown2013,Popova2013}. Improved technologies and survey
strategies allow for the discovery of more and smaller NEOs than ever
before. However, resources for NEO characterization lag behind and are
usually limited to the study of the brightest and hence usually
largest NEOs. This lack in physical and compositional data compromises
the predictions of current NEO distribution models
\citep[e.g.,][]{Bottke2002}, which assume a uniform and
size-independent compositional distribution throughout the entire NEO
population. Studying the physical properties of NEOs allows us to test
this assertion and provide important constraints for future NEO
distribution models. Furthermore, the comparison of the compositional
distribution of NEOs with those of meteorite falls provides clues on
asteroid strengths and is key to properly assess the threat to Earth
through future asteroid impacts.

A common way to investigate asteroid compositions is through
spectroscopy. Spectroscopic observations allow for the identification
of both the overall continuum shape and diagnostic band features,
enabling their classification into different taxonomies. Some asteroid
taxonomic types can be related to meteorites to understand detailed
composition. In this work, we make use of the widely used Bus-DeMeo
taxonomy scheme \citep{DeMeo2009}, which combines optical with
near-infrared (NIR) spectra, covering the wavelength range
0.45--2.45~$\mu\mathrm{m}$. Most asteroids observed so far can be
classified into one of 3 major complexes: silicaceous S-type
asteroids, carbonaceous C-type asteroids, and the X-type complex. 
  Taxonomic complexes are sets of taxonomic types with similar
  spectral properties. However, not all taxonomic types are part of a
  complex; some taxonomic types, e.g. V-type and D-type asteroids,
  have spectra that are very distinct from those of other types and
  complexes. C-type and X-type asteroids have very similar, feature-less
spectra, which makes it hard to distinguish between the two.

The most commonly used instrument/telescope combinations in asteroid
studies \citep[e.g., NASA's InfraRed Telescope Facility with its SpeX
spectrograph,][]{Rayner2003} have effective limiting magnitudes around
$V \sim$~18.0. Specialized characterization surveys like the {\it
  Mission Accessible Near-Earth Objects Survey}
\citep[MANOS,][]{Moskovitz2015} are able to extend this spectroscopic
coverage to $V \leq$ 21 for a small sample of 10-15 NEOs on
Earth-like, mission-accessible orbits that can be observed each
month. For comparison, current asteroid discovery surveys, e.g., the
Catalina Sky Survey and PanSTARRS-1, discover on average 2 NEOs per
night, most of which are in the brightness range $19 < V < 21$ at the
time of discovery \citep{Galache2015}.  The limited sensitivity of
spectroscopic surveys in most cases forces a lower limit on the sizes
of asteroids that can be observed and characterized.  In order to
increase the fraction of characterized NEOs and provide a more
homogeneous characterization as a function of asteroid size, more
telescope time and/or a more efficient observing approach is
necessary.

Asteroid taxonomic classification relies on low-resolution reflectance
spectra. In order to estimate the spectral type of a NEO, photometric
measurements at a few key wavelengths --- a method referred to
as {\it spectrophotometry} --- are usually
sufficient. Spectrophotometry has the advantage of being more
sensitive in terms of target brightness because the light is collected
within a bandpass instead of being dispersed as a function of
wavelength. Spectrophotometric observations have been
used in the past to classify asteroid taxonomies, including the
eight-color asteroid survey \citep{Zellner1985}, the 52-color asteroid
survey \citep{Bell2005}, the Sloan Digital Sky Survey
\citep[e.g.,][]{GilHutton2010}, and the 2MASS Asteroid and Comet
Survey \citep{Sykes2000}. Here we present a new approach in which we
combine spectrophotometry with {\it rapid response observations},
i.e., observations that are obtained shortly after the discovery of
the target, in order to observe and characterize even small NEOs with
a higher efficiency than current spectroscopic methods \citep[see][for
a discussion]{Galache2015}.

\begin{figure}
 \centering
 \includegraphics[width=\linewidth]{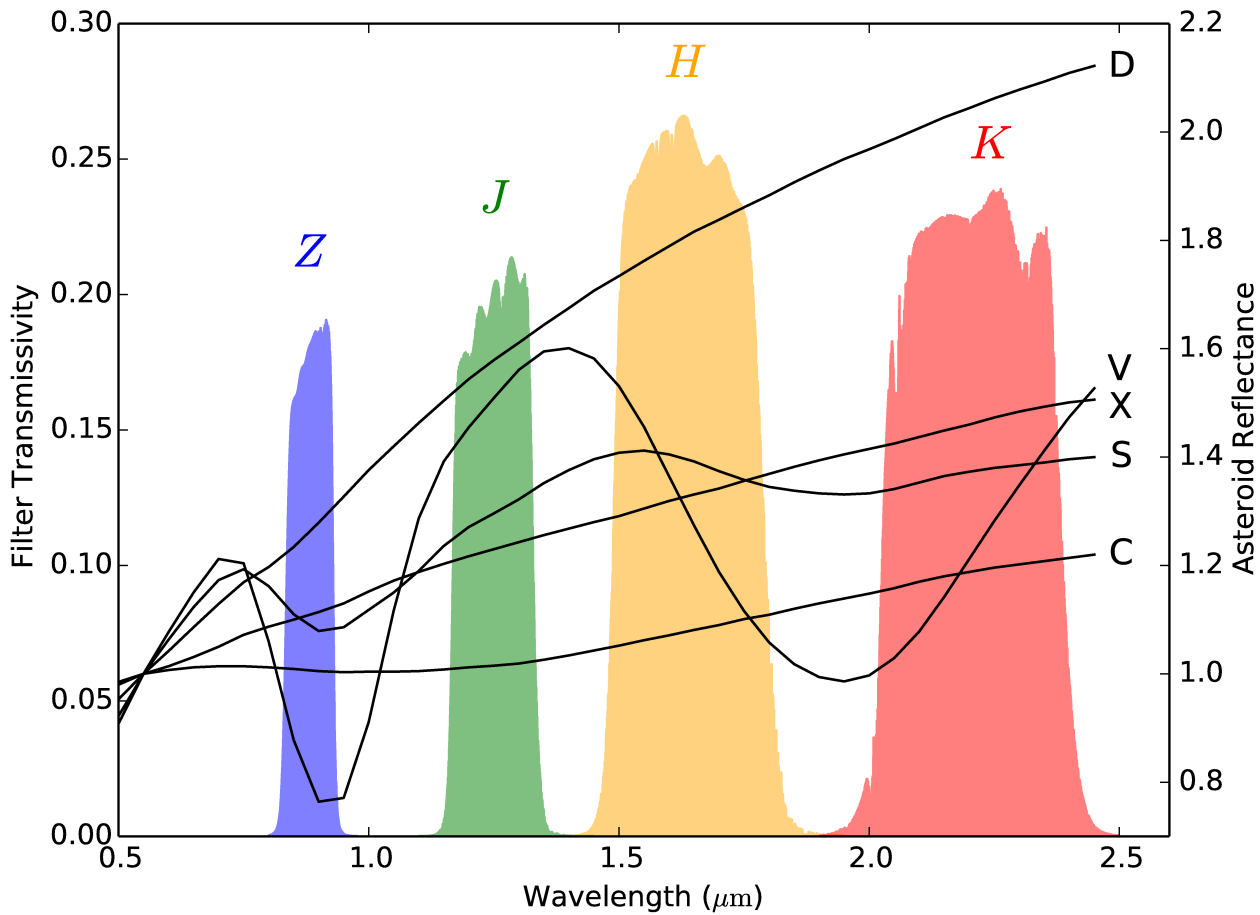}
 \caption{NIR photometric colors are indicative of asteroid
   compositions. The $Z$, $J$, $H$, and $K$ bandpasses used by UKIRT
   \citep{Hewett2006} are plotted together with averaged asteroid
   reflectance spectra \citep[black lines,][]{DeMeo2009} of the most
   common asteroid taxonomic types and complexes (sets of
     taxonomic types with similar spectral properties.)}
 \label{fig:filters_spectra}
\end{figure}

\section{Observations and Data Analysis}
\label{lbl:observations}

Observations for this project are performed with the {\it Wide Field
  Camera} for UKIRT \citep[WFCAM,][]{Casali2007}. The {\it
  United Kingdom Infrared Telescope} (UKIRT), which is now operated by
the University of Hawaii, the University of Arizona, and the Lockheed
Martin Advanced Technology Center, is a 3.8~m Cassegrain-type
telescope located on Maunakea, Hawai'i. WFCAM consists of four
2k$\times$2k detectors, each of which covers a field of the sky with
an edge length of 13.7\arcmin. The WFCAM photometric system
\citep{Hewett2006} is mostly identical to the wide-spread Mauna Kea
Observatory $J$, $H$, and $K$ near-infrared filters and includes a $Z$
band filter that is very similar to the SDSS $z$ band filter. All
these filters have been used in the {\it UKIRT Infrared Deep Sky
  Survey} \citep[UKIDSS,][]{Lawrence2007}. UKIRT is operated in queue
mode, allowing for flexibility in the scheduling of observations. We
started observations for this project in semester 2014A and obtained
110 observations of 104 different NEOs in semesters 2014A and
2014B. Observations are still ongoing as of writing this.

\subsection{Observation Planning}

In this program, we acquire observations in the $Z$, $J$, $H$, and $K$
bands.  $H$ is the only filter that has not been used for all targets;
we added $H$ later to compensate for a potential loss of $Z$ data due
to calibration issues.  Figure \ref{fig:filters_spectra} shows that
these filter bands sample the spectral slope and silicate band
features of the most common asteroid taxonomies well. Our approach is
to measure the brightness of our target NEOs in these bands and then
compare the differential colors (e.g., $Z-J$, $J-K$) to color data
synthesized from measured asteroid spectra (Section
\ref{lbl:synthesis}).  Our {\it rapid response} approach is key to the
success of our program. Most NEOs are discovered when they are
brightest, i.e., closest to the observer. After their discovery, NEOs
fade quickly at a rate of typically 0.5~mag within one week and 5~mag
within 6 weeks as they increase their distance from the observer
\citep{Galache2015}. By triggering rapid response spectrophotometric
observations of NEOs within a few days of discovery, we are able to
observe and characterize objects with absolute magnitudes $H \sim 28$,
i.e., with diameters of a few meters. Such rapid response is generally
not feasible through classical observing proposals to heavily
oversubscribed major research facilities.

Potential targets are identified and uploaded into the {\it UKIRT}
queue on a daily basis. Accessible targets are identified among those
NEOs that have been discovered within the last 4 weeks; this duration
is somewhat arbitrary, but usually leads to a number of
well-observable and bright potential targets. Since most targets fade
quickly in brightness, those that get observed were discovered only a
few days before their observation. In the case of the unavailability
of rapid-response targets, we upload other accessible NEOs into the
queue, which we refer to as ``substitute'' targets. A target is
considered accessible if it has a visible brightness $V \leq 22$, and
an airmass $\leq$ 2.0, as provided by the JPL Horizons system
\citep{Giorgini1996}, for at least the duration of the estimated
integration time from Maunakea. Potential targets are manually
selected from the list of accessible targets, prioritizing objects
with high absolute magnitudes $H_V$ (small sizes) and large values of
$H_V-V$, where $V$ is the apparent magnitude of the target in the coming
night. High values of $H_V-V$ ensure that our targets are observed when
they are close to the Earth.  {\it UKIRT} queue observing scripts are
automatically created for the selected targets, using the latest
orbital elements of the targets as provided by JPL Horizons. The
telescope tracks the motion of the target, leading to a trailing of
the background sources. We use a frame time of 5~s in each band to
minimize the trailing; most targets move at ${\leq}5$~arcsec per
minute. The total integration time in each band is a function of the
predicted target brightness $V$, typically varying between 30~min and
2~hr in all bands. In our observations, we place the target in the
center of {\it WFCAM} camera 3, which provides the best noise
properties. Different combinations of dither patterns (3.2\arcsec\
step size) and additional telescope offsets are used to enable the
creation of proper flatfield images from the imaging data that are used
to mitigate against pixel-to-pixel variability in the detector. In
order to account for rotational brightness changes in our targets we
intersperse $Z$, $H$, and $K$ observations with $J$ observations, in
which the targets are usually bright. A typical filter sequence for a
bright target is $JKJZJHJKJ$; fainter targets require longer
integration times, involving repetitions of this pattern. Typical
frame numbers per filter [$J$, $Z$, $H$, $K$] are [50, 40, 40, 70] for
bright targets ($V\leq19.5$) and [110, 50, 50, 320] for faint targets
($V>20.5$); the integration time is 5~s for each frame. The larger
  number of frames in $K$ band is a result of the degraded detector
  sensitivity, the higher background, and reduced solar emission in
  this band.

\subsection{Data Analysis}

Basic data reduction is performed by the Cambridge Astronomical Survey
Unit (CASU) 
using the default {\it UKIRT} data reduction recipes
\citep{Irwin2006}.  Reduced individual frames are downloaded from
CASU, registered based on 2MASS \citep{Skrutskie2006} catalog stars in
the field, and combined in the frame of the sky (``{\it skycoadd}''
frame) and in the moving frame of the target (``{\it comove}'' frame)
using {\tt SOURCE EXTRACTOR}, {\tt SCAMP}, and {\tt SWARP}
\citep{Bertin1996, Bertin2002, Bertin2006}. Individual frames are
combined in groups of observations in the same band that were taken
consecutively (one {\it member} of the filter sequence), retaining the
filter sequence and providing a skycoadd and comove frame for each
filter sequence member. All comove images are inspected to make sure
that the target is not contaminated from background sources; if
necessary, individual contaminated frames are excluded from the
combination process to mitigate against contamination.  Aperture
photometry is obtained using {\tt SOURCE EXTRACTOR}. The optimum
aperture size is derived from the comove and skycoadd images of each
filter sequence member using a curve-of-growth method
\citep{Howell2006}.  A common aperture radius is used for all
observations of one target that is based on the fraction of flux
enclosed for the target and background sources, as well as the
resulting signal-to-noise ratios.  In the rare case of strongly
trailed background sources, we manually select a larger aperture
radius to assure that similar flux ratios for both the target and the
background sources are enclosed.
The magnitude zeropoint of each filter sequence member is derived from
the skycoadd image, using an uncertainty-weighted
$\chi^2$-minimization based on available 2MASS field stars. Typical
zeropoint uncertainties are of the order of 0.01~mag or less, using
catalog magnitudes from ${\sim}$100 2MASS stars converted into the
UKIRT photometric system \citep{Hodgkin2009}. 

In order to check the consistency of our photometric calibration, we
compare magnitudes measured with our pipeline with standard star
magnitudes from the literature for select fields. We obtain reduced
image data of 5 standard star fields \citep{Leggett2006} from the {\it
  WFCAM Science Archive}\footnote{{\tt http://surveys.roe.ac.uk/wsa/}}
in the $J$, $H$, and $K$ bands and process them using our default
analysis pipeline. We find average magnitude offsets between our
2MASS-calibrated measurements and the literature magnitudes of the
order of 0.03~mag, which is smaller than the typical photometric
uncertainties observed in our targets.

\label{lbl:zband_calibration}
$Z$ band calibration requires additional treatment, as it is subject
to non-negligible reddening effects due to galactic extinction. The
transformation from 2MASS $J$, $H$, and $Ks$ (a short-bandpass version
of the common $K$ band filter), all of which are barely impacted by
reddening, into the UKIRT $Z$ band, which is subject to reddening,
requires an offset that was derived within this work.
We compare {\it Sloan Digital Sky Survey} (SDSS, data release 9) $z$
band magnitudes with $Z$ band magnitudes transformed from 2MASS into
UKIRT magnitudes for different fields in the sky as a function of
galactic latitude and longitude. We find the offset between both
catalogs to be stable at (0.064$\pm$0.035)~mag with respect to
galactic longitude and latitudes $|b| > 15$\degr, but highly variable
for $|b| < 15$\degr. 
Where available, we hence obtain our $Z$ band calibration from SDSS
$z$ band data, which accounts for reddening. Alternatively, we convert
2MASS data into $Z$ band using the transformation given by
\citet{Hodgkin2009} and add the offset derived above. We consider $Z$
band calibrations based on 2MASS data at $|b| < 15$\degr\ unreliable
and flag targets accordingly (see Section \ref{lbl:results}).


From the 110 observations of 104 different NEOs we obtained in
2014A/B, we had to reject 18 observations mostly because they were too
faint and the chosen exposure time was too short or the background was
too crowded. 
We adjusted the total integration time for subsequent observations so
that our final sample includes 92 useful observations of 86 NEOs.

\subsection{Lightcurve Variability Correction and Color Measurement}
\label{lbl:lightcurve}

\begin{figure}
 \centering
 \includegraphics[width=\linewidth]{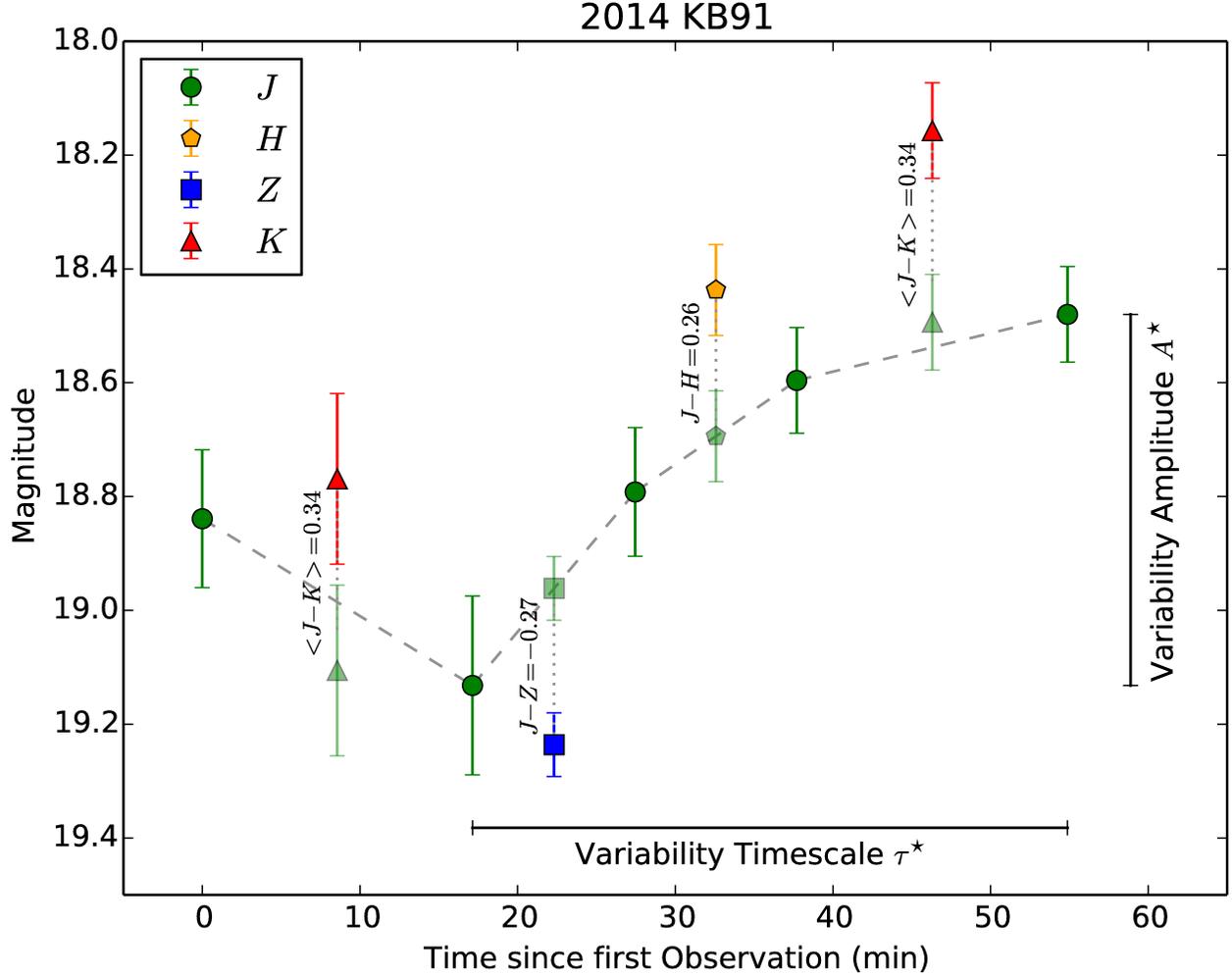}
 \caption{Color determination for NEO 2014~KB91. The target's
     lightcurve is clearly visible in the $J$ band observations (green
     circles). We interpolate between $J$ band observations to obtain
     the lightcurve-corrected brightness in $J$ at times of non-$J$
     observations (dashed line and semi-transparent symbols, see
     text). Colors are derived by subtracting the non-$J$ magnitudes
     from these interpolated $J$ magnitudes. The measured colors for
     2014~KB91 are $J-H$=-0.05$\pm$0.17, $Z-J$=-0.09$\pm$0.20,
     $J-K$=-0.02$\pm$0.15, after subtraction of the solar colors. Note
     that the derived $J-K$ color is a weighted average of two
     observations; hence, the interpolated $J$ magnitudes for the $K$
     band are slightly offset from the linear interpolation, but still
     agree within 1$\sigma$. We also derive the variability amplitude
     ($A^\star=0.65\pm0.18$) and the variability timescale
     ($\tau^\star\geq37.7$~min) from our data (see Section
     \ref{lbl:lightcurve} for details).}
 \label{fig:lightcurve_example}
\end{figure}

Irregularly shaped NEOs exhibit potentially significant brightness
variations as a function of time. In order to account for the
asteroids' lightcurves, we intersperse our observations in the
different filter bands with $J$ band observations. Most targets
exhibit a clear variability in the $J$ band. Note that this
variability is not caused by transparency variations, which are
accounted for through the absolute photometric calibration of our
target fields. Aiming for an accurate measurement of the target's
  color, we have to account for the target's brightness variability
  that is inherent to our brightness measurements. We base this
  correction on the interspersed $J$ band observations.  For each
[$H$, $K$, $Z$] band observation, we derive the lightcurve-corrected
$J$ band magnitude at the time of the observation through linear
interpolation based on those $J$ band observations that are closest in
time before and after the non-$J$ band observation. By
  interpolating the $J$ band brightness at the times of non-$J$
  observations, we obtain an approximate measure for the brightness in
  $J$ at these time. Based on the interpolated $J$ band magnitude, we
derive the corresponding colors $J-$[$H$, $K$, $Z$] (see Figure
\ref{fig:lightcurve_example}).  Using the interpolated $J$ band
  brightness, we can derive the target's color more accurately than by
  simply subtracting $J$ from non-$J$ measurements.  The uncertainty
of each color measurement is defined as the quadratic sum of the
uncertainties of all three observations (2$\times J$ plus one non-$J$
band magnitude uncertainty), which results from Gaussian error
  propagation. If more than one observation is available in either
non-$J$ band, the corresponding color is derived as the weighted mean
over all measurements of that color, where the weight is the inverse
squared color uncertainty. The total color uncertainty is the
quadratic sum of the root-sum-square of the involved magnitude
uncertainties and the standard deviation of the mean of the individual
color determinations, if several measurements are available.  From the
measured colors we subtract the colors of the Sun, which we obtained
with the synthesis method introduced in Section \ref{lbl:synthesis}.

As a by-product, our method also allows for a qualitative description
of the target's lightcurve. Although our observations usually only
cover a fraction of the rotational period of each target, we are able
to put constraints on both the target's brightness variability
amplitude ($A^\star$) and timescale ($\tau^\star$). We define the
variability amplitude as the maximum difference in $J$ band magnitudes
(also including interpolated $J$ magnitudes from non-$J$ measurements)
and the timescale as the difference in time between those
measurements. If a lightcurve measurement exhibits more than one
maximum and minimum, we measure the amplitude and timescale between
two consecutive extrema. The method is graphically presented in Figure
\ref{fig:lightcurve_example}. Note that $A^\star$ and $\tau^\star$
should not be mistaken as the target's rotational amplitude and
period. $A^\star$ generally provides a lower limit on the lightcurve
amplitude of the target as we have to assume that only part of the
lightcurve has been sampled; its uncertainty is derived as the
quadratic sum of the magnitude uncertainties of the maximum and
minimum brightness, providing a measure for the significance of
$A^\star$. $\tau^\star$ only provides a measure for the temporal
variability of the brightness of the target; it may suggest a fast or
slowly rotating nature of the target.

\section{Synthesis of NEO NIR Colors from Measured Spectra}
\label{lbl:synthesis}

\begin{figure}
 \centering
 \includegraphics[width=0.7\linewidth]{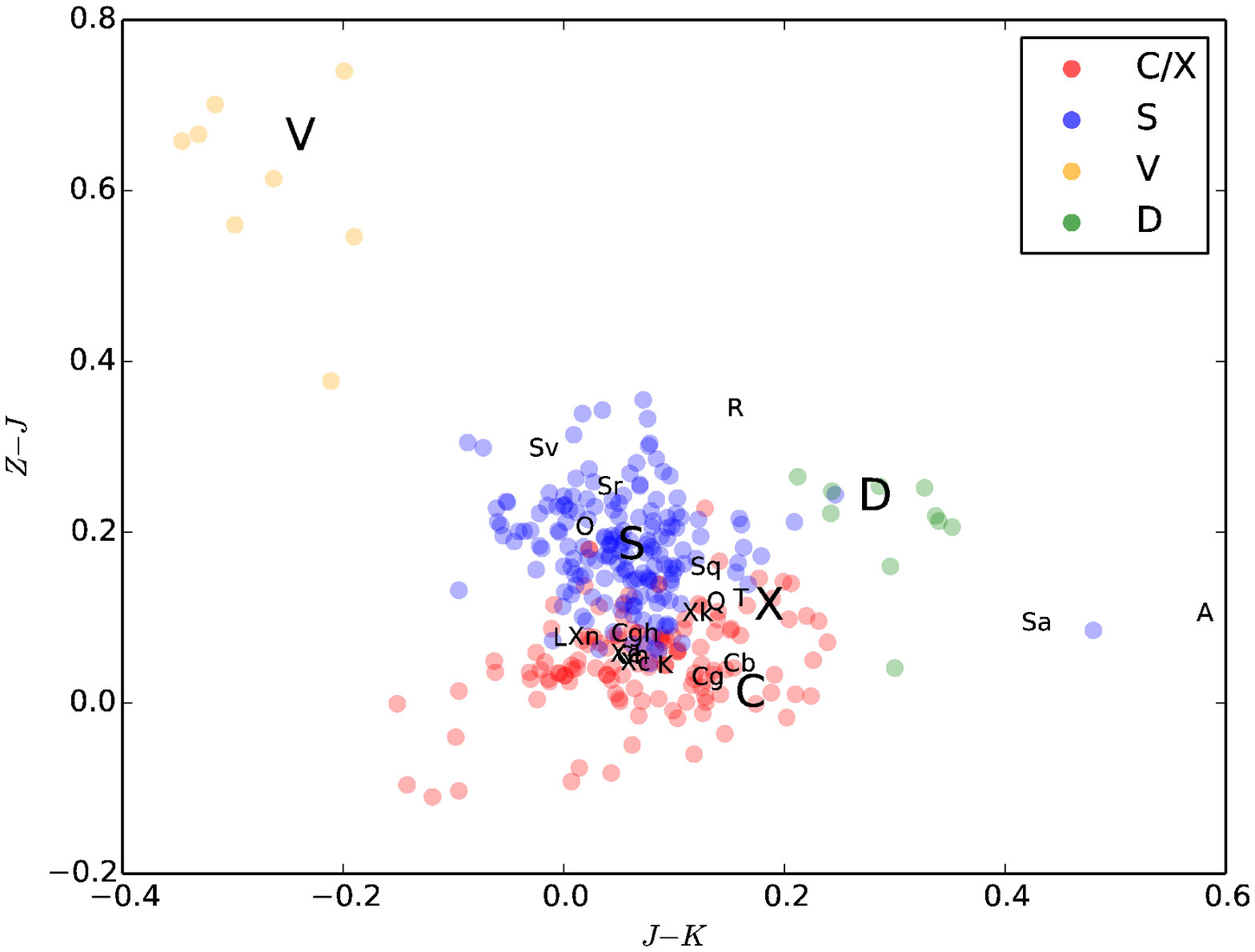}
 \vspace{2em}\par
 \includegraphics[width=0.7\linewidth]{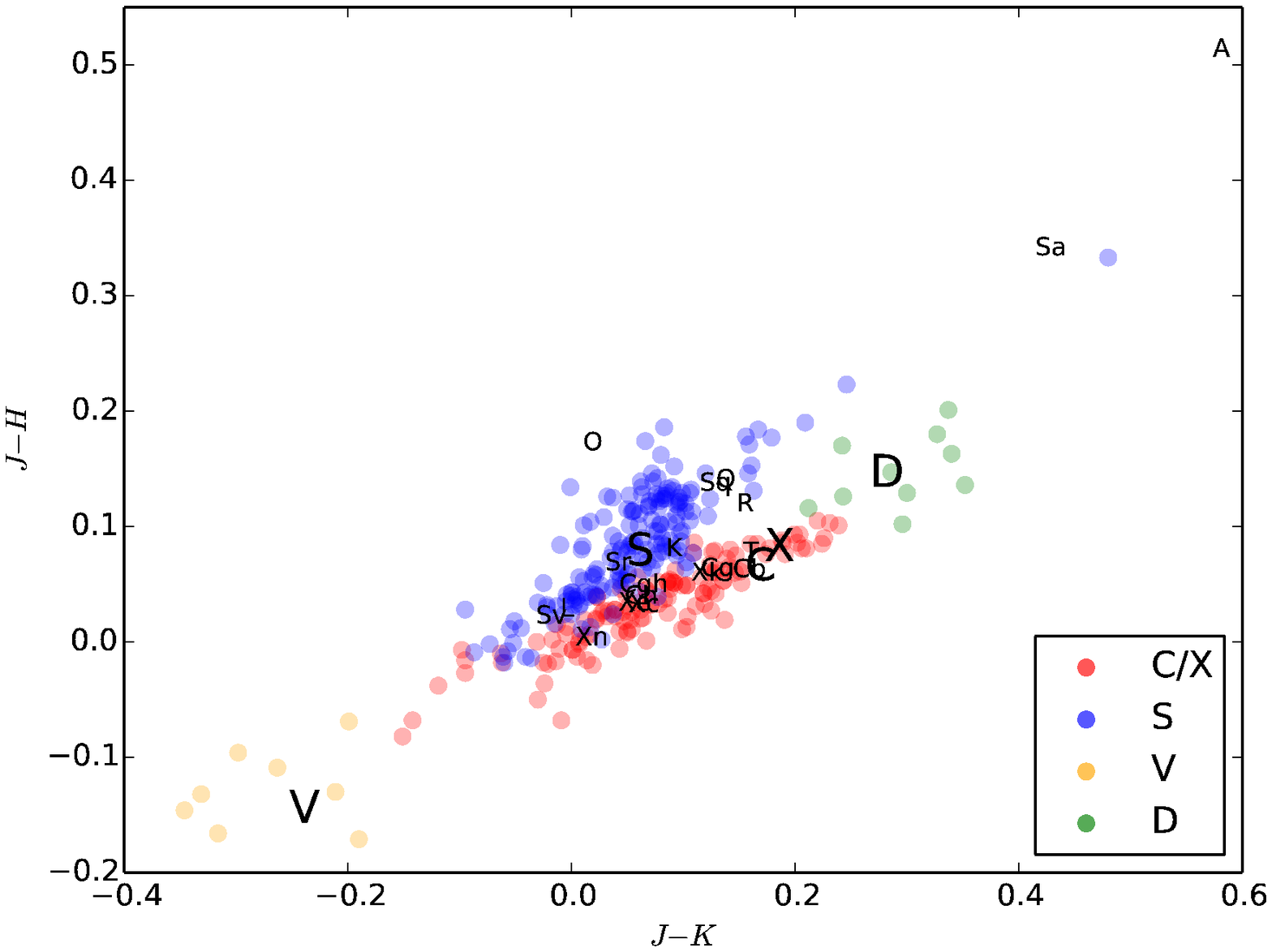}
 \caption{Asteroid color training sample derived from MIT-UH-IRTF
   data. Letters indicate the average locations of individual
   taxonomic types; taxonomic complexes and types that can be
   identified in this work are highlighted \citep[average reflectance
   spectra from][]{DeMeo2009}. }
 \label{fig:colorspace}
\end{figure}

In order to interpret our measured asteroid colors, we synthesize
asteroid colors from measured asteroid spectra. The largest available
database of NEO spectra is the MIT-UH-IRTF Joint Campaign for NEO
Spectral Reconnaissance\footnote{{\tt
    http://smass.mit.edu/minus.html}}, which obtains NIR spectra of
NEOs using {\it SpeX} \citep{Rayner2003} on NASA's Infrared Telescope
Facility. We obtained 614 asteroid spectra from that survey (including
a number of duplicate observations) and classified them using the
Bus-DeMeo taxonomy spectrum classification on-line
routine\footnote{{\tt http://smass.mit.edu/busdemeoclass.html}}. We
only used spectra that cover the wavelength range from 0.8 to
2.45~$\mu$m and accepted only those classifications from the on-line
routine with absolute average residuals of 0.05 or less. We also
allowed for multiple classifications if the residuals were within 10\%
of the minimum residual value. While these restrictions reduce the
number of classified spectra (439 spectra), they drastically increase
the quality of the synthesized colors. The color-color distribution of
different taxonomic types in the Bus-DeMeo System is shown in Figure
\ref{fig:colorspace}.  The plots show that spectroscopic differences
between taxonomic complex sub-types and even some independent
taxonomic types are subtle. Hence, we decided to distinguish only
between a few distinct complexes and types: S-complex (including 
  types S, Sa, Sq, Sqw, Sr, Srw, Sv, and Sw; 171 objects),
C-complex and X-complex (including types C, Cb, Cg,
Cgh, Ch, X, Xc, Xe, Xk, and Xn; 130 objects), and a small number of
independent taxonomic types (D-type and V-type, 18 objects in
total). Note that the spectral similarity between C-complex and
X-complex asteroids complicates a distinction between members of these
complexes. This simplified classification scheme \citep[compared to
the full Bus-DeMeo scheme,][]{DeMeo2009} implies the possibility of
mis-classifications due to a lack of less frequent asteroid taxonomic
types in our scheme. Figure \ref{fig:colorspace} shows that many less
frequent asteroid types fall within the color-space occupied by the
main complexes: A, R, O, Q, and L-type asteroids are likely to be
mistaken for S-complex asteroids, whereas T and K-type asteroids might
be classified as C/X-complex asteroids. This ambiguity within the
smaller taxonomic type groups, as well as the ambiguity between the C
and X-complexes could only be resolved with a significantly higher
precision in the measurement of the asteroid colors.

Each classified spectrum is turned into synthesized colors in the
UKIRT photometric bands $Z$, $J$, $H$, and $K$ \citep[bandpass
information is provided by][]{Hewett2006}. In order to derive the
absolute flux $F_{f,t}$ in a certain bandpass, we convolve the
spectral response $R_{f}(\lambda)$ of each filter $f$ with the product
of the respective asteroid reflectance spectrum $A_t(\lambda)$ of
taxonomic type $t$ and a spectrum of the Sun\footnote{{\tt
    http://kurucz.harvard.edu/stars/sun/fsunallp.10000resam25}}
$S(\lambda)$ over the whole bandpass wavelength range:

\begin{equation}
  F_{f,t} = \int R_{f}(\lambda) \; A_t(\lambda) S(\lambda) \; \lambda \ d\lambda.
\label{eqn:convolve}
\end{equation}

From this flux, we derive the resulting synthetic magnitude and
calibrate it in the Vega magnitude system. We derive the zeropoint
magnitude using Equation \ref{eqn:convolve} by setting
$A_t(\lambda)\equiv1$ and replacing $S(\lambda)$ with a spectrum of
Vega\footnote{{\tt
    http://kurucz.harvard.edu/stars/vega/vegallpr25.10000resam25}}. Colors
are derived from the calibrated magnitudes and solar colors
($J-K=0.354$, $Z-J=0.369$, $J-H=0.304$) are subtracted.
Color-color plots are displayed in Figure
\ref{fig:colorspace}, showing the distribution of synthesized asteroid
spectra of different taxonomic types in $Z-J$ vs.\ $J-K$ and $J-H$
vs.\ $J-K$ color spaces.

\section{Taxonomic Classification}
\label{lbl:classification}

We classify our sample targets based on the sample of synthesized
asteroid NIR colors (see Section \ref{lbl:synthesis}) in a
machine-learning approach using the {\tt scikit-learn} module
\citep{Pedregosa2011} for {\tt Python}. In order to find the most
reliable classification algorithm for our problem, we tested the
accuracy of different methods, including different Nearest-Neighbor,
Support Vector Machine (SVM), and a Gaussian Naive Bayes method, that
are provided within {\tt scikit-learn} (see {\tt
  http://scikit-learn.org} for discussions of the individual
methods). In our test, we remove one object at a time from the sample
of 319 synthesized asteroid NIR colors of S-complex, C/X-complex,
D-type, and V-type asteroids (see Section \ref{lbl:synthesis}) and
predict its class based on the different methods that have been
trained on the rest of the sample. Table \ref{tbl:classification}
compares the numbers of correct classifications for the different
methods, considering two different cases: the first case assumes two
color measurements ($Z-J$ and $J-K$); the second case assumes three
color measurements ($Z-J$, $J-H$, and $J-K$).  The accuracy of the
algorithms varies between the two cases; generally, the 3-color case
derives more accurate results based on the fact that the
increased number of degrees of freedom puts additional constraints on
the classification problem. In the 2-color case, the $k$-Nearest
Neighbor ($k=5$) and the Gaussian Naive Bayes methods achieve the same
overall accuracy; in the 3-color case, the $k$-Nearest Neighbor
($k=1$) is the most accurate method. Hence, we adopt the $k$-Nearest
Neighbor ($k=1$) method if 3 asteroid colors are available and the
Gaussian Naive Bayes method if only 2 colors are available. We refrain
from using $k$-Nearest Neighbor ($k=5$) as it relies on fairly large
training samples for the different taxonomic types, which are not
available in the case of V-type and D-type asteroids.

\begin{deluxetable}{rcc}
\tablewidth{0pt}
\tablecaption{Classification Algorithm Accuracy Comparison.\label{tbl:classification}}
\tablehead{
\colhead{Algorithm} &
\colhead{2 Colors} &
\colhead{3 Colors}}
\startdata

Nearest Centroid & 85.6\% & 89.3\%\\
$k$-Nearest Neighbor ($k$=1) & 85.9\% & {\bf 96.2\%}\\
$k$-Nearest Neighbor ($k$=3) & 87.8\% & 94.4\%\\
$k$-Nearest Neighbor ($k$=5) & {\bf 88.4\%} & 95.0\%\\
SVM, linear kernel & 85.0\% & 91.2\%\\
SVM, polynomial kernel (3rd degree) & 53.6\% & 53.6\%\\
SVM, radial basis function kernel & 85.0\% & 91.2\%\\
Gaussian Naive Bayes &  {\bf 88.4\%} & 92.5\%

\enddata
\end{deluxetable}

In order to account for uncertainties in the measurement of each
target's color, we use a Monte Carlo approach in which we randomize
the measured colors within the corresponding 1$\sigma$ uncertainties
in each color based on a Gaussian distribution in $10^6$ trials. We
classify the whole ensemble using the respective algorithm trained on
our set of synthesized asteroid NIR colors and count the frequency of
classifications in the individual taxonomic types. Thus, we obtain
classification probabilities for the individual taxonomies and we
adopt the most likely taxonomic class for the target.

\section{Results}
\label{lbl:results}

\begin{figure}
 \centering
 \includegraphics[width=\linewidth]{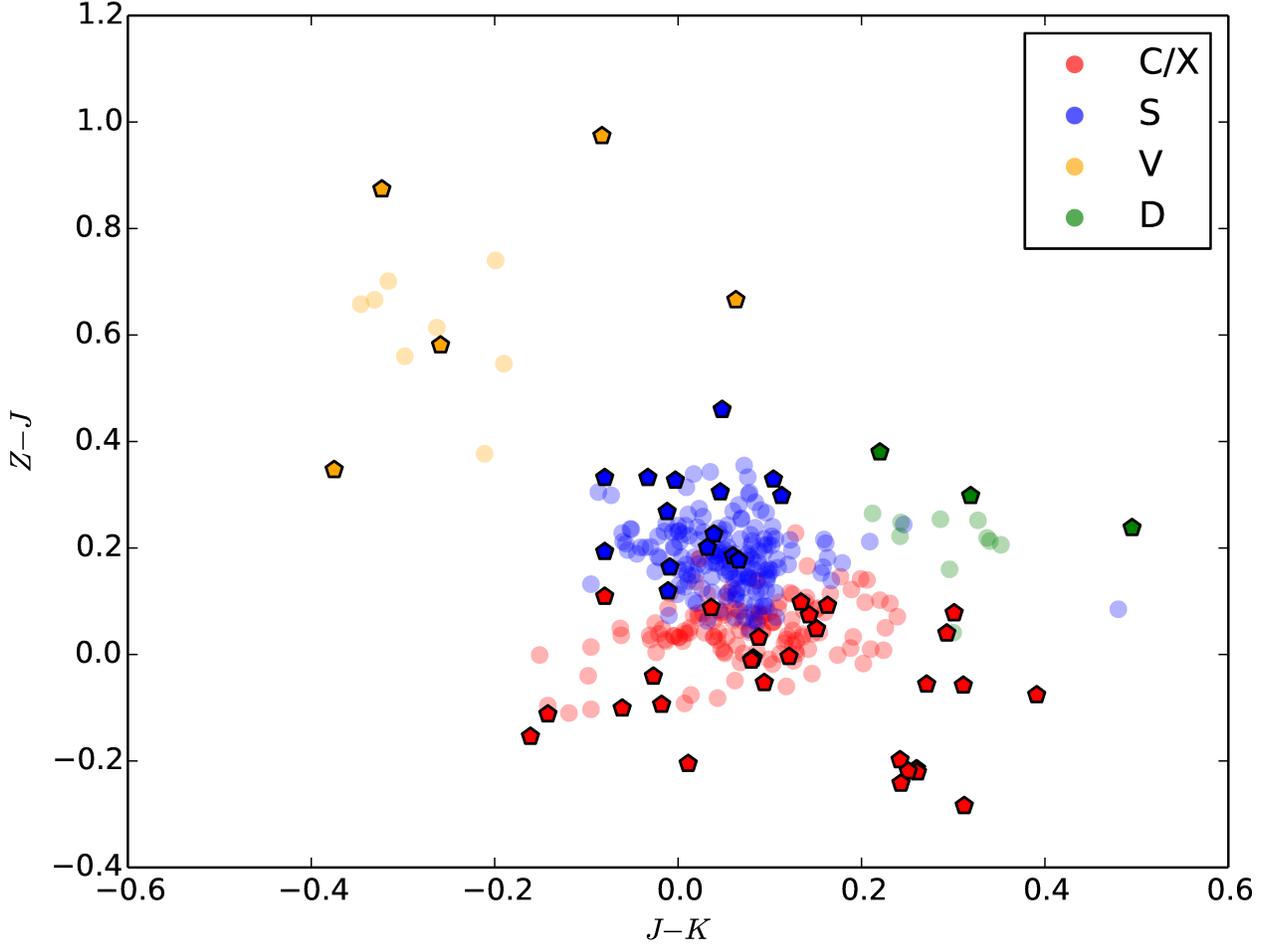}
 \caption{Distribution of observed NEOs in $Z-J$ and $J-K$ color
   space. Semi-transparent circles represent the training samples shown
   in Figure \ref{fig:colorspace}. The color of each pentagon reflects
   the most likely taxonomic classification based on the Monte Carlo
   method described in Section \ref{lbl:classification} using the
   measured $Z-J$ and $J-K$ colors only. In this plot, we only show
   NEOs with a minimum probability of 50\% for a taxonomic class, a
   root-mean-square color uncertainty of less than 0.3~mag, and
   observations that are not affected by irregularities (51 out of 67
   targets with $Z$, $J$, $K$ observations meet these criteria, see
   Table \ref{tbl:results}). Solar colors have been subtracted from
   each datapoint. }
 \label{fig:trainingsample_results}
\end{figure}

Figure \ref{fig:trainingsample_results} shows an example color-color
plot that is based on measured $Z-J$ and $J-K$ colors of our sample
targets. The probabilistic classification results for all data are
presented in Table \ref{tbl:results}. Some of our sample targets were
observed at low galactic latitudes and calibrated using $Z$ band
magnitudes that were transformed from 2MASS data, potentially leading
to inaccuracies in the $Z$ magnitudes (see Section
\ref{lbl:zband_calibration}). Furthermore, some targets exhibit highly
variable lightcurves that make it impossible to put constraints on the
variability timescale, or show potential inconsistencies in the
measured lightcurve. Both effects can potentially influence the
reliability of the color measurements (see Section
\ref{lbl:discussion} for a discussion). Taxonomic classifications that
are subject to either of the irregularities listed above should be
considered with care; affected targets are marked in the Notes column
of Table \ref{tbl:results}.

Only considering classifications with a probability ${\geq}50$\%,
total root-mean-square color uncertainties ${\leq}0.3$~mag, and
observations that are not affected by irregularities as discussed
above leads to reliable classifications for 46 observations of 43
different NEOs. 25 more observations are subject to irregularities, 18
observations suffer from a low signal-to-noise ratio (color
uncertainties ${>}0.3$~mag), and 7 observations lead to ambiguous
taxonomic classifications (probabilities for all taxonomic
classifications ${<}50$\%). Note that the ratio of reliable
classifications is higher for those targets that were observed more
recently, which is a result of an increase of integration time in our
observation planning.

\begin{deluxetable}{rccccccccccccccl}
\tabletypesize{\tiny}
\rotate
\tablewidth{0pt}
\tablecaption{Observations and Results.\label{tbl:results}}
\tablehead{
\colhead{Object} &
\colhead{Obs. Midtime} &
\colhead{Dur.} &
\colhead{$H_V$} &
\colhead{$V$} &
\colhead{$J_{\mathrm{Median}}$} &
\colhead{$Z-J$} &
\colhead{$J-H$} &
\colhead{$J-K$} &
\colhead{$A^\star$} &
\colhead{$\tau^\star$} &
\colhead{C/X} &
\colhead{D} &
\colhead{S} &
\colhead{V} &
\colhead{Notes} \\
\colhead{} &
\colhead{(UT)} &
\colhead{(hr)} &
\colhead{(mag)} &
\colhead{(mag)} &
\colhead{(mag)} &
\colhead{(mag)} &
\colhead{(mag)} &
\colhead{(mag)} &
\colhead{(mag) } &
\colhead{(min)} &
\colhead{Prob.} &
\colhead{Prob.} &
\colhead{Prob.} &
\colhead{Prob.} &
\colhead{}}
\startdata
       99799$^\star$ & 14-03-21 08:51 & 0.3 & 18.30 & 19.89 & 18.91 &  0.81$\pm$0.41 & \nodata &  0.21$\pm$0.31 & 0.40$\pm$0.27 & ${\geq}$7.0 & 0.06 & 0.29 & 0.08 & 0.57 & \\
      125268$^\star$ & 15-01-18 06:32 & 2.0 & 16.10 & 20.06 & 18.59 &  0.30$\pm$0.11 & \nodata &  0.11$\pm$0.08 & 0.20$\pm$0.17 &  40.0 & 0.05 & 0.09 & {\bf 0.84} & 0.01& \\
      141857$^\star$ & 15-01-17 07:12 & 2.0 & 16.30 & 19.83 & 18.31 &  0.52$\pm$0.12 & \nodata & -0.04$\pm$0.07 & 0.60$\pm$0.10 &  36.6 & 0.00 & 0.00 & 0.33 & {\bf 0.67}& \tablenotemark{1} \\
   2004~JN13$^\star$ & 15-01-15 07:00 & 0.6 & 15.30 & 16.55 & 15.12 & -0.07$\pm$0.02 & \nodata &  0.11$\pm$0.01 & 0.05$\pm$0.01 & ${\geq}$33.6 & {\bf 1.00} & 0.00 & 0.00 & 0.00& \tablenotemark{1} \\
     2006~WW$^\star$ & 15-01-17 14:16 & 1.9 & 16.10 & 19.90 & 18.67 &  0.54$\pm$0.10 & \nodata &  0.17$\pm$0.08 & 0.27$\pm$0.12 & ${\geq}$44.2 & 0.00 & 0.41 & 0.43 & 0.16& \\
   2009~TG10$^\star$ & 14-04-15 11:40 & 0.2 & 17.60 & 16.90 & 15.28 & -0.11$\pm$0.02 & \nodata & -0.14$\pm$0.04 & 0.09$\pm$0.05 & ${\geq}$6.5 & {\bf 1.00} & 0.00 & 0.00 & 0.00& \\
   2010~AG79$^\star$ & 14-12-29 06:32 & 1.5 & 19.90 & 20.86 & 19.74 & -0.53$\pm$0.27 & \nodata &  0.67$\pm$0.32 & 0.66$\pm$0.47 &  22.0 & 0.97 & 0.00 & 0.00 & 0.03& \tablenotemark{1} \tablenotemark{2} \\
   2011~CH50$^\star$ & 14-04-04 05:33 & 0.2 & 21.80 & 18.48 & 17.48 & -0.06$\pm$0.10 & \nodata &  0.31$\pm$0.08 & 0.17$\pm$0.10 &   4.0 & {\bf 0.90} & 0.10 & 0.00 & 0.00& \\
   2011~CH50$^\star$ & 14-04-16 09:57 & 0.2 & 21.80 & 18.16 & 16.99 & -0.00$\pm$0.07 & \nodata &  0.12$\pm$0.10 & 0.13$\pm$0.16 &   5.0 & {\bf 0.94} & 0.01 & 0.05 & 0.00& \\
    2011~OL5$^\star$ & 14-12-29 15:35 & 0.6 & 20.20 & 19.28 & 18.50 &  0.18$\pm$0.13 & \nodata &  0.09$\pm$0.19 & 0.36$\pm$0.29 &  19.5 & 0.30 & 0.19 & 0.47 & 0.04& \\
   2011~WK15$^\star$ & 14-12-29 14:58 & 0.6 & 19.70 & 18.29 & 16.90 &  0.05$\pm$0.04 & \nodata &  0.15$\pm$0.04 & 0.08$\pm$0.03 & ${\geq}$33.8 & {\bf 0.96} & 0.00 & 0.04 & 0.00& \\
   2012~CO46$^\star$ & 14-12-29 11:53 & 1.4 & 22.80 & 20.26 & 19.41 & -0.06$\pm$0.25 & \nodata &  0.27$\pm$0.18 & 0.52$\pm$0.27 & ${\geq}$44.2 & 0.73 & 0.17 & 0.09 & 0.00& \\
   2013~BM76$^\star$ & 15-01-17 15:49 & 0.9 & 20.20 & 19.76 & 18.80 &  0.04$\pm$0.14 & \nodata &  0.29$\pm$0.11 & 0.10$\pm$0.16 & ${\geq}$42.4 & {\bf 0.63} & 0.30 & 0.07 & 0.00& \\
   2013~BM76$^\star$ & 15-01-20 15:31 & 0.9 & 20.20 & 19.74 & 18.76 &  0.33$\pm$0.12 & \nodata &  0.10$\pm$0.09 & 0.19$\pm$0.09 & ${\geq}$23.1 & 0.04 & 0.10 & {\bf 0.83} & 0.03& \\
    2014~GF1 & 14-04-09 10:39 & 1.1 & 26.00 & 21.21 & 19.85 &  0.24$\pm$0.25 & \nodata &  0.49$\pm$0.18 & 0.11$\pm$0.22 & ${\geq}$37.4 & 0.38 & {\bf 0.58} & 0.03 & 0.01& \\
   2014~GH17 & 14-04-09 09:08 & 1.3 & 21.90 & 20.33 & 19.47 &  0.30$\pm$0.25 & \nodata &  0.32$\pm$0.23 & 0.17$\pm$0.22 &  12.9 & 0.25 & {\bf 0.50} & 0.19 & 0.06& \\
   2014~GV48 & 14-04-13 08:25 & 1.1 & 25.80 & 20.18 & 19.15 &  0.28$\pm$0.27 & \nodata &  0.04$\pm$0.16 & 0.59$\pm$0.22 & ${\geq}$29.5 & 0.27 & 0.11 & 0.41 & 0.21& \\
    2014~HE3 & 14-10-12 10:23 & 0.5 & 19.90 & 19.09 & 18.07 & -0.27$\pm$0.13 &  0.24$\pm$0.16 &  0.13$\pm$0.15 & 0.34$\pm$0.14 &  10.4 & {\bf 0.73} & 0.04 & 0.23 & 0.00& \tablenotemark{2} \\
     2014~HW & 14-04-24 11:50 & 2.2 & 28.40 & 18.94 & 17.67 &  0.33$\pm$0.11 & \nodata &  0.06$\pm$0.08 & 0.40$\pm$0.20 &  12.0 & 0.02 & 0.02 & {\bf 0.91} & 0.04& \tablenotemark{2} \\
   2014~JA31 & 14-05-12 13:51 & 0.4 & 23.60 & 19.18 & 18.21 & -0.04$\pm$0.17 & \nodata & -0.03$\pm$0.20 & 0.30$\pm$0.18 & ${\geq}$ & {\bf 0.82} & 0.02 & 0.12 & 0.03& \\
   2014~JG78 & 14-05-21 12:15 & 0.8 & 19.60 & 20.44 & 19.75 & -0.17$\pm$0.31 & \nodata &  0.27$\pm$0.32 & 0.70$\pm$0.57 & ${\geq}$25.3 & 0.80 & 0.10 & 0.06 & 0.05& \\
   2014~JR25 & 14-05-11 15:00 & 0.4 & 23.40 & 18.39 & 17.20 & -0.20$\pm$0.09 & \nodata &  0.01$\pm$0.09 & 0.20$\pm$0.07 & ${\geq}$12.3 & {\bf 1.00} & 0.00 & 0.00 & 0.00& \\
   2014~JS57 & 14-05-18 13:51 & 0.8 & 22.20 & 20.42 & 18.84 & -0.06$\pm$0.16 & \nodata &  0.07$\pm$0.23 & 0.55$\pm$0.38 & ${\geq}$25.0 & {\bf 0.85} & 0.05 & 0.09 & 0.02& \tablenotemark{2} \\
   2014~JT54 & 14-05-12 08:45 & 0.9 & 24.10 & 20.19 & 18.82 &  0.26$\pm$0.21 & \nodata & -0.08$\pm$0.18 & 0.56$\pm$0.23 & ${\geq}$35.6 & 0.26 & 0.04 & 0.40 & 0.30& \\
   2014~JV54 & 14-05-20 14:01 & 0.4 & 25.90 & 20.33 & 19.16 &  0.14$\pm$0.30 & \nodata & -0.08$\pm$0.27 & 0.56$\pm$0.28 & ${\geq}$22.0 & 0.45 & 0.07 & 0.21 & 0.26& \\
   2014~JW24 & 14-05-11 06:36 & 0.4 & 23.90 & 19.77 & 18.60 &  0.22$\pm$0.26 & \nodata & -0.10$\pm$0.20 & 0.46$\pm$0.25 & ${\geq}$21.8 & 0.35 & 0.04 & 0.30 & 0.31& \\
   2014~JY30 & 14-05-11 07:01 & 0.4 & 25.40 & 20.12 & 19.09 &  0.32$\pm$0.41 & \nodata & -0.49$\pm$0.32 & 0.59$\pm$0.38 &   3.2 & 0.23 & 0.01 & 0.05 & 0.71& \\
   2014~KA46 & 14-06-08 10:19 & 0.9 & 24.40 & 20.09 & 18.58 &  0.28$\pm$0.21 & -0.36$\pm$0.18 &  0.08$\pm$0.25 & 0.94$\pm$0.23 & ${\geq}$17.4 & 0.40 & 0.10 & 0.13 & 0.37& \\
   2014~KB91 & 14-06-04 11:10 & 0.9 & 20.20 & 19.88 & 18.79 & -0.09$\pm$0.20 & -0.05$\pm$0.17 & -0.02$\pm$0.15 & 0.65$\pm$0.18 & ${\geq}$37.7 & {\bf 0.83} & 0.01 & 0.15 & 0.02& \\
     2014~KD & 14-05-20 13:36 & 0.4 & 24.40 & 18.95 & 16.63 &  0.21$\pm$0.08 & \nodata &  0.17$\pm$0.10 & 0.53$\pm$0.12 &   9.0 & 0.11 & 0.28 & {\bf 0.61} & 0.00& \tablenotemark{2} \\
   2014~KO62 & 14-06-03 12:07 & 1.0 & 26.20 & 20.29 & 18.97 &  0.38$\pm$0.15 & -0.18$\pm$0.22 &  0.22$\pm$0.18 & 0.97$\pm$0.19 &  30.0 & 0.15 & 0.29 & 0.38 & 0.18& \\
   2014~KX86 & 14-06-03 11:02 & 0.9 & 26.50 & 21.03 & 19.37 &  0.87$\pm$0.22 & \nodata & -0.32$\pm$0.33 & 1.16$\pm$0.36 & ${\geq}$27.6 & 0.00 & 0.03 & 0.01 & 0.96& \\
    2014~MC6 & 14-10-11 11:10 & 1.0 & 19.40 & 19.62 & 18.74 & -0.08$\pm$0.18 &  0.10$\pm$0.17 &  0.39$\pm$0.16 & 0.39$\pm$0.18 & ${\geq}$27.8 & 0.34 & 0.40 & 0.26 & 0.00& \\
   2014~ME18 & 14-07-26 08:03 & 1.4 & 23.00 & 21.08 & 19.85 &  0.15$\pm$0.26 &  0.31$\pm$0.29 &  0.08$\pm$0.27 & 0.80$\pm$0.46 &  24.4 & 0.17 & 0.10 & 0.68 & 0.04& \tablenotemark{1} \\
   2014~MJ27 & 14-07-26 09:42 & 1.5 & 23.10 & 20.91 & 19.89 &  0.08$\pm$0.33 & -0.22$\pm$0.38 &  0.30$\pm$0.26 & 0.92$\pm$0.58 &  17.9 & 0.47 & 0.24 & 0.19 & 0.09& \\
   2014~MK55 & 14-12-30 09:50 & 1.1 & 21.40 & 20.40 & 19.39 &  0.36$\pm$0.23 & \nodata &  0.11$\pm$0.36 & 0.47$\pm$0.57 &  32.8 & 0.17 & 0.30 & 0.24 & 0.29& \\
    2014~MK6 & 14-07-25 07:47 & 1.0 & 21.00 & 19.61 & 18.87 & -0.21$\pm$0.26 &  0.47$\pm$0.27 &  0.26$\pm$0.22 & 0.40$\pm$0.34 & ${\geq}$17.4 & 0.25 & 0.06 & 0.69 & 0.00& \\
   2014~MM55 & 14-10-13 13:24 & 0.5 & 18.70 & 18.79 & 17.78 & -0.22$\pm$0.11 &  0.07$\pm$0.11 &  0.26$\pm$0.08 & 0.42$\pm$0.09 & ${\geq}$29.6 & {\bf 0.91} & 0.08 & 0.01 & 0.00& \\
   2014~MP41 & 14-07-12 13:50 & 0.9 & 18.40 & 19.73 & 19.61 & -0.16$\pm$0.33 &  0.16$\pm$0.32 & -0.04$\pm$0.27 & 0.57$\pm$0.27 & ${\geq}$38.7 & 0.59 & 0.04 & 0.34 & 0.04& \\
   2014~MQ60 & 14-07-25 09:51 & 1.3 & 22.40 & 20.44 & 19.22 &  0.58$\pm$0.26 &  0.18$\pm$0.31 & -0.26$\pm$0.28 & 0.85$\pm$0.32 &  12.0 & 0.03 & 0.01 & 0.32 & 0.64& \\
     2014~MX & 14-07-25 11:38 & 1.4 & 24.10 & 20.84 & 19.78 & -0.20$\pm$0.25 & -0.27$\pm$0.69 &  0.24$\pm$0.29 & 1.71$\pm$0.76 &  16.1 & 0.67 & 0.08 & 0.22 & 0.03& \\
   2014~MX17 & 14-07-11 11:14 & 0.9 & 20.50 & 19.62 & 18.38 &  0.11$\pm$0.10 &  0.07$\pm$0.10 &  0.04$\pm$0.09 & 0.16$\pm$0.09 & ${\geq}$18.0 & 0.41 & 0.01 & {\bf 0.58} & 0.00& \tablenotemark{1} \\
   2014~NB52 & 14-07-11 13:09 & 1.4 & 21.70 & 20.18 & 19.32 & -0.15$\pm$0.27 &  0.38$\pm$0.26 &  0.34$\pm$0.18 & 0.69$\pm$0.26 & ${\geq}$33.3 & 0.25 & 0.13 & 0.62 & 0.00& \tablenotemark{1} \tablenotemark{3} \\
   2014~NC39 & 14-07-16 09:58 & 0.9 & 21.70 & 20.54 & 19.85 & -0.18$\pm$0.37 & \nodata &  0.57$\pm$0.30 & 0.55$\pm$0.43 & ${\geq}$55.1 & 0.84 & 0.12 & 0.02 & 0.02& \\
   2014~ND52 & 14-07-12 12:22 & 1.4 & 23.70 & 20.33 & 19.24 &  0.29$\pm$0.19 & -0.04$\pm$0.23 &  0.14$\pm$0.17 & 0.50$\pm$0.23 & ${\geq}$33.7 & 0.24 & 0.19 & 0.46 & 0.11& \\
    2014~NE3 & 14-07-11 12:04 & 0.5 & 20.10 & 18.58 & 17.49 & -0.01$\pm$0.09 &  0.12$\pm$0.09 &  0.08$\pm$0.07 & 0.28$\pm$0.09 & ${\geq}$ & {\bf 0.56} & 0.01 & 0.43 & 0.00& \\
   2014~NE39 & 14-07-18 10:47 & 0.9 & 20.60 & 20.15 & 18.99 & -0.05$\pm$0.16 &  0.17$\pm$0.17 &  0.09$\pm$0.16 & 0.31$\pm$0.16 & ${\geq}$19.0 & {\bf 0.50} & 0.08 & 0.43 & 0.00& \\
    2014~NG3 & 14-07-10 14:26 & 0.9 & 24.20 & 20.06 & 19.02 & -0.28$\pm$0.18 &  0.54$\pm$0.19 &  0.31$\pm$0.13 & 0.41$\pm$0.17 &  10.3 & 0.17 & 0.03 & {\bf 0.79} & 0.00& \\
   2014~NL52 & 14-07-15 11:58 & 1.4 & 23.60 & 20.34 & 19.63 & \nodata &  0.23$\pm$0.21 &  0.37$\pm$0.23 & 0.61$\pm$0.29 &  25.2 & 0.25 & 0.30 & 0.45 & 0.00& \tablenotemark{3} \\
    2014~OG1 & 14-07-25 08:37 & 0.5 & 21.60 & 18.48 & 17.57 & -0.22$\pm$0.17 &  0.06$\pm$0.14 &  0.25$\pm$0.13 & 0.57$\pm$0.15 & ${\geq}$ & {\bf 0.81} & 0.14 & 0.06 & 0.00& \\
  2014~OT111 & 14-07-31 09:44 & 0.5 & 21.70 & 17.39 & 15.92 &  0.07$\pm$0.03 &  0.11$\pm$0.03 &  0.14$\pm$0.02 & 0.10$\pm$0.02 & ${\geq}$14.4 & {\bf 0.60} & 0.00 & 0.40 & 0.00& \\
    2014~OV3 & 14-07-29 09:26 & 0.9 & 23.20 & 19.43 & 17.88 &  0.67$\pm$0.09 & -0.07$\pm$0.09 &  0.06$\pm$0.08 & 0.43$\pm$0.09 & ${\geq}$27.5 & 0.00 & 0.00 & 0.25 & {\bf 0.75}& \\
    2014~OV3 & 15-01-18 13:14 & 1.4 & 23.20 & 20.41 & 19.18 &  0.20$\pm$0.20 & \nodata &  0.18$\pm$0.18 & 0.60$\pm$0.18 & ${\geq}$79.0 & 0.31 & 0.31 & 0.34 & 0.04& \tablenotemark{2} \\
   2014~PL51 & 14-10-11 10:10 & 0.5 & 20.40 & 18.84 & 17.61 &  0.10$\pm$0.11 &  0.20$\pm$0.12 & -0.01$\pm$0.12 & 0.30$\pm$0.12 & ${\geq}$19.6 & 0.15 & 0.01 & {\bf 0.83} & 0.00& \tablenotemark{1} \\
   2014~RC12 & 14-10-12 15:19 & 1.0 & 18.80 & 19.86 & 18.48 &  0.06$\pm$0.14 &  0.32$\pm$0.14 & -0.10$\pm$0.11 & 0.58$\pm$0.13 & ${\geq}$30.0 & 0.11 & 0.00 & {\bf 0.89} & 0.00& \tablenotemark{1} \\
   2014~RL12 & 15-01-18 11:26 & 1.9 & 17.90 & 17.93 & 16.66 &  0.23$\pm$0.03 & \nodata &  0.03$\pm$0.05 & 0.19$\pm$0.08 &  89.3 & 0.00 & 0.00 & {\bf 1.00} & 0.00& \tablenotemark{1} \\
   2014~RL12 & 15-01-19 07:38 & 2.1 & 17.90 & 18.00 & 16.61 &  0.16$\pm$0.03 & \nodata & -0.07$\pm$0.02 & 0.11$\pm$0.03 & ${\geq}$99.3 & 0.02 & 0.00 & {\bf 0.98} & 0.00& \tablenotemark{1} \\
   2014~RQ17 & 14-12-29 13:36 & 1.9 & 22.30 & 20.70 & 19.47 &  0.77$\pm$0.22 & \nodata &  0.05$\pm$0.19 & 0.52$\pm$0.29 &  11.4 & 0.00 & 0.14 & 0.09 & {\bf 0.77}& \tablenotemark{3} \\
  2014~SF145 & 14-12-29 10:47 & 0.6 & 22.60 & 19.31 & 18.10 &  0.12$\pm$0.11 & \nodata &  0.15$\pm$0.10 & 0.17$\pm$0.14 &  12.2 & 0.48 & 0.13 & 0.38 & 0.00& \tablenotemark{1} \\
  2014~SM143 & 14-10-12 14:30 & 0.5 & 20.30 & 17.15 & 15.97 &  0.09$\pm$0.03 &  0.21$\pm$0.03 &  0.16$\pm$0.02 & 0.06$\pm$0.03 & ${\geq}$28.9 & 0.02 & 0.00 & {\bf 0.98} & 0.00& \\
  2014~SM261 & 14-10-12 11:16 & 0.5 & 21.10 & 19.05 & 17.87 &  0.10$\pm$0.13 &  0.22$\pm$0.14 &  0.31$\pm$0.11 & 0.19$\pm$0.14 & ${\geq}$14.3 & 0.13 & 0.43 & 0.44 & 0.00& \\
    2014~SS1 & 14-10-12 09:36 & 0.5 & 21.70 & 18.75 & 17.49 &  0.18$\pm$0.10 &  0.14$\pm$0.10 &  0.06$\pm$0.08 & 0.30$\pm$0.09 & ${\geq}$10.0 & 0.14 & 0.02 & {\bf 0.84} & 0.00& \\
  2014~SZ264 & 14-10-11 12:10 & 0.9 & 18.60 & 19.98 & 18.68 &  0.10$\pm$0.14 &  0.29$\pm$0.16 &  0.34$\pm$0.13 & 0.13$\pm$0.15 &  20.7 & 0.08 & 0.32 & {\bf 0.60} & 0.00& \\
   2014~TJ17 & 14-10-11 08:58 & 0.5 & 24.90 & 18.70 & 17.64 &  0.33$\pm$0.14 &  0.06$\pm$0.14 & -0.00$\pm$0.15 & 0.40$\pm$0.13 & ${\geq}$7.0 & 0.09 & 0.05 & {\bf 0.75} & 0.11& \\
   2014~TN17 & 14-10-10 07:12 & 0.5 & 21.50 & 19.07 & 17.72 &  0.24$\pm$0.14 & -0.10$\pm$0.13 & -0.00$\pm$0.24 & 0.55$\pm$0.26 &   9.9 & 0.31 & 0.12 & 0.36 & 0.21& \tablenotemark{2} \tablenotemark{3} \\
     2014~TV & 14-10-11 09:36 & 0.5 & 24.40 & 18.91 & 18.04 & -0.24$\pm$0.26 &  0.61$\pm$0.24 &  0.24$\pm$0.16 & 0.82$\pm$0.25 & ${\geq}$14.8 & 0.16 & 0.02 & 0.82 & 0.00& \\
   2014~TZ17 & 14-10-16 14:40 & 0.9 & 22.70 & 19.62 & 18.21 &  0.12$\pm$0.09 &  0.08$\pm$0.09 & -0.01$\pm$0.09 & 0.22$\pm$0.08 & ${\geq}$38.1 & 0.29 & 0.00 & {\bf 0.71} & 0.00& \\
  2014~UA176 & 14-11-05 11:19 & 0.6 & 26.70 & 18.94 & 17.69 &  0.30$\pm$0.10 & \nodata &  0.05$\pm$0.08 & 0.16$\pm$0.12 & ${\geq}$26.4 & 0.03 & 0.02 & {\bf 0.92} & 0.03& \\
  2014~UF206 & 15-01-15 07:51 & 0.8 & 18.80 & 15.41 & 13.90 & -0.01$\pm$0.01 & \nodata &  0.08$\pm$0.01 & 0.10$\pm$0.01 & ${\geq}$33.8 & {\bf 1.00} & 0.00 & 0.00 & 0.00& \\
     2014~US & 14-11-05 10:38 & 0.6 & 19.10 & 18.14 & 16.84 &  0.11$\pm$0.05 & \nodata & -0.08$\pm$0.03 & 0.04$\pm$0.03 & ${\geq}$14.9 & {\bf 0.54} & 0.00 & 0.46 & 0.00& \\
   2014~UT33 & 14-11-04 14:10 & 0.6 & 23.40 & 18.37 & 17.10 & -0.25$\pm$0.06 & \nodata &  0.01$\pm$0.05 & 0.05$\pm$0.05 & ${\geq}$14.3 & {\bf 1.00} & 0.00 & 0.00 & 0.00& \tablenotemark{1} \\
  2014~UV115 & 14-11-05 14:11 & 0.6 & 22.70 & 19.17 & 17.50 &  0.21$\pm$0.06 & \nodata &  0.20$\pm$0.04 & 0.15$\pm$0.05 &   5.1 & 0.10 & 0.27 & {\bf 0.63} & 0.00& \tablenotemark{1} \\
  2014~UV210 & 15-01-17 12:15 & 1.9 & 26.90 & 20.65 & 19.65 & -0.25$\pm$0.30 & \nodata &  0.25$\pm$0.31 & 1.00$\pm$0.31 &  23.1 & 0.85 & 0.06 & 0.04 & 0.05& \tablenotemark{3} \\
     2014~VM & 14-11-10 11:51 & 0.6 & 17.70 & 17.97 & 16.66 &  0.33$\pm$0.04 & \nodata & -0.08$\pm$0.06 & 0.09$\pm$0.03 & ${\geq}$19.5 & 0.00 & 0.00 & {\bf 0.89} & 0.11& \\
     2014~VP & 14-11-08 11:34 & 0.9 & 22.80 & 19.51 & 18.23 &  0.03$\pm$0.12 & \nodata & -0.21$\pm$0.11 & 0.16$\pm$0.10 & ${\geq}$22.9 & {\bf 0.79} & 0.00 & 0.12 & 0.09& \tablenotemark{1} \\
  2014~WJ201 & 14-12-07 12:06 & 0.9 & 24.80 & 19.84 & 19.28 &  0.35$\pm$0.27 & \nodata & -0.38$\pm$0.25 & 0.23$\pm$0.22 &   7.9 & 0.16 & 0.01 & 0.10 & 0.73& \\
   2014~WJ70 & 15-01-16 14:10 & 1.4 & 17.70 & 19.25 & 18.05 &  0.10$\pm$0.08 & \nodata &  0.13$\pm$0.06 & 0.18$\pm$0.07 & ${\geq}$23.0 & {\bf 0.59} & 0.02 & 0.39 & 0.00& \\
    2014~WP4 & 14-12-02 12:24 & 1.1 & 24.30 & 20.01 & 18.55 &  0.97$\pm$0.15 & \nodata & -0.08$\pm$0.09 & 0.16$\pm$0.10 & ${\geq}$38.7 & 0.00 & 0.00 & 0.00 & {\bf 1.00}& \\
    2014~XJ3 & 14-12-15 10:09 & 0.6 & 20.10 & 17.16 & 15.70 &  0.27$\pm$0.03 & \nodata & -0.01$\pm$0.03 & 0.07$\pm$0.02 & ${\geq}$19.9 & 0.00 & 0.00 & {\bf 1.00} & 0.00& \\
     2014~YD & 15-01-15 13:13 & 1.4 & 24.30 & 20.14 & 18.78 &  0.16$\pm$0.18 & \nodata & -0.01$\pm$0.12 & 0.40$\pm$0.15 &  22.0 & 0.39 & 0.02 & {\bf 0.51} & 0.07& \\
   2014~YE35 & 15-01-13 09:25 & 0.6 & 20.30 & 18.49 & 17.11 &  0.46$\pm$0.06 & \nodata &  0.05$\pm$0.06 & 0.10$\pm$0.05 & ${\geq}$19.4 & 0.00 & 0.00 & {\bf 0.88} & 0.12& \\
   2014~YU41 & 15-01-14 13:14 & 2.0 & 23.80 & 21.00 & 19.94 & -0.02$\pm$0.26 & \nodata &  0.23$\pm$0.19 & 0.56$\pm$0.41 & ${\geq}$62.1 & 0.68 & 0.18 & 0.13 & 0.01& \tablenotemark{3} \\
   2014~YV34 & 15-01-13 08:42 & 0.6 & 19.50 & 18.55 & 17.26 &  0.23$\pm$0.06 & \nodata &  0.04$\pm$0.04 & 0.07$\pm$0.05 & ${\geq}$14.4 & 0.02 & 0.00 & {\bf 0.98} & 0.00& \\
   2014~YW34 & 15-01-16 11:07 & 1.4 & 21.60 & 20.46 & 19.27 &  0.18$\pm$0.21 & \nodata &  0.07$\pm$0.12 & 0.48$\pm$0.17 &  22.8 & 0.38 & 0.07 & {\bf 0.50} & 0.05& \\
   2014~YY43 & 15-01-18 14:29 & 0.9 & 19.40 & 19.71 & 18.77 &  0.19$\pm$0.14 & \nodata & -0.08$\pm$0.10 & 0.32$\pm$0.11 &   7.9 & 0.30 & 0.00 & {\bf 0.58} & 0.11& \\
   2015~AN44 & 15-01-20 10:34 & 1.4 & 25.20 & 20.38 & 19.17 &  0.20$\pm$0.14 & \nodata &  0.03$\pm$0.12 & 0.57$\pm$0.15 & ${\geq}$33.2 & 0.27 & 0.04 & {\bf 0.65} & 0.04& \\
   2015~AP44 & 15-01-20 12:23 & 1.9 & 26.20 & 20.45 & 19.54 &  0.03$\pm$0.20 & \nodata &  0.09$\pm$0.12 & 0.48$\pm$0.19 & ${\geq}$88.7 & {\bf 0.65} & 0.05 & 0.28 & 0.01& \\
   2015~AR45 & 15-01-20 09:27 & 0.6 & 19.80 & 19.32 & 18.12 &  0.33$\pm$0.11 & \nodata & -0.03$\pm$0.14 & 0.29$\pm$0.22 &  19.3 & 0.03 & 0.04 & {\bf 0.70} & 0.23& \\
     2015~BD & 15-01-20 13:51 & 0.6 & 23.90 & 19.51 & 18.44 &  0.15$\pm$0.12 & \nodata &  0.23$\pm$0.14 & 0.51$\pm$0.08 & ${\geq}$19.6 & 0.35 & 0.36 & 0.29 & 0.00& \tablenotemark{2} \\
   2015~BG92 & 15-01-24 12:16 & 0.4 & 25.10 & 19.08 & 17.80 &  0.09$\pm$0.10 & \nodata &  0.04$\pm$0.10 & 0.11$\pm$0.08 & ${\geq}$14.5 & {\bf 0.59} & 0.01 & 0.39 & 0.00& \\
   2015~BG92 & 15-01-25 09:50 & 0.4 & 25.10 & 18.92 & 17.52 & -0.15$\pm$0.10 & \nodata & -0.16$\pm$0.10 & 0.32$\pm$0.09 & ${\geq}$19.8 & {\bf 1.00} & 0.00 & 0.00 & 0.00& \\
   2015~BG92 & 15-01-26 09:59 & 0.6 & 25.10 & 18.75 & 17.46 & -0.10$\pm$0.06 & \nodata & -0.06$\pm$0.05 & 0.23$\pm$0.05 & ${\geq}$14.4 & {\bf 1.00} & 0.00 & 0.00 & 0.00& \\

\enddata
\tablenotetext{\star}{{Substitute} target (see Section \ref{lbl:observations});}
\tablenotetext{1}{{low} galactic latitude (target was observed at $b
  \leq 15$\degr) and $Z$ band calibration from transformed 2MASS data;}
\tablenotetext{2}{{lightcurve} inconsistencies in at least one band
  (non-$J$ measurements do not agree with $J$-lightcurve trend);}
\tablenotetext{3}{highly variable lightcurve (no reliable
  determination of $\tau^\star$ possible).}

\tablecomments{The table lists for each target its official number or
  designation, observation midtime of the entire filter sequence, the
  observation duration, the absolute ($H_V$) and apparent visible
  ($V$) magnitude as provided by JPL Horizons at the time of the
  observation, the median $J$ band brightness from all our
  measurements, the measured color indices (from which solar colors
  have been subtracted), the variability amplitude ($A^\star$) and
  timescale ($\tau^\star$), and the classification probabilities for
  the individual taxonomic complexes and types. For each target we use
  all available colors ($Z-J$, $J-K$, and $J-H$, where available) to
  provide the best-possible accuracy in our taxonomic classification;
  bold numbers highlight reliable classifications with a probability
  ${\geq}50$\% and total root-mean-square color uncertainties
  ${\leq}0.3$~mag.  Note that faint targets usually have high
  uncertainties in their color measurements, leading to a less
  reliable classification result.}

\end{deluxetable}


\section{Discussion}
\label{lbl:discussion}

We investigate the consistency of our color measurements and
classifications based on three targets with two or more observations
that do not suffer from irregularities (notes in Table
\ref{tbl:results}): 2011~CH50, 2013~BM76, and 2015~BG92. These targets
were observed twice or more as a result of an aborted
observation due to rapidly changing weather conditions and hence have
shorter than intended integration times. Duplicate observations were
also obtained in a few cases by the {\it UKIRT} telescope operators to
test telescope operations. We find multiple color measurements and
variability amplitudes of all objects to agree within 1--2$\sigma$. 
In Table \ref{tbl:results}, we marked sample targets
that show suspicious lightcurve behavior in the form of lightcurve
inconsistencies (marked with ``2'' in the Notes column) or a highly
variable lightcurve (marked with ``3'' in the Notes column).
Nevertheless, we believe that our linear-interpolation approach
provides the most robust results in the majority of cases. Non-linear
interpolation, e.g., using third-order polynomials or splines, would
require additional lightcurve information that is currently not
available for our sample targets. In a few cases, discrepancies in
$Z-J$ color measurement might also be caused by the lack of extinction
correction in the transformation of 2MASS data into $Z$ band data (see
Section \ref{lbl:zband_calibration}; targets that are potentially
affected by this effect are marked with ``1'' in the Notes columns of
Table \ref{tbl:results}).

We compare our results to optical spectra obtained within the MANOS
project \citep[MANOS classifications taken from][]{Hinkle2015}; we
could not find spectral data of our sample targets in the further
literature. We find an overlap of 2 targets with MANOS: 2014~HW
(MANOS: S-type, this work: 91\% S-complex probability) and 2014~UV210
(MANOS: Cb/Cgh-type, this work: 85\% C-complex probability). Note that
both targets suffer from irregularities (notes in Table
\ref{tbl:results}) and we are still able to reproduce the MANOS
results with a high level of confidence.  Five of our sample targets
have measured lightcurves, which allows us to compare our variability
amplitude ($A^\star$) and timescale ($\tau^\star$) to the measured
rotational amplitude and period. \citet{Warner2015b} finds a period of
31.3~min and an amplitude of 0.99~mag for 2014~RQ17, for which we find
$\tau^\star=11.4$~min and $A^\star=0.52\pm0.29$~mag. For 2014~SS1,
\citet{Warner2015a} finds a period of 16.63~hr and an amplitude of
0.43~mag; we find $\tau^\star\geq10$~min and
$A^\star=0.30\pm0.09$~mag. Additional information is available
from the MANOS project for 2014~HW (3.8~min, 1.07~mag, our results:
$\tau^\star=12$~min, $A^\star=0.4\pm0.2$~mag), 2014~UV210 (33.4~min,
0.91~mag, our results: $\tau^\star=23.1$~min,
$A^\star=1.00\pm0.31$~mag), 2015~BG92 (10.7~min, 0.36~mag, our
results: $\tau^\star\geq14.4$~min, $A^\star\geq0.11$~mag). For 
  three out of five targets with lightcurve data we find $\tau^\star$
to be smaller than the rotational period, which is a result of the
fact that $\tau^\star$ only provides a sense of the timescale on which
the lightcurve changes. In those two cases in which $\tau^\star$
  is greater than the rotational period, we account this to the fact
  that our sampling frequency is too coarse to properly resolve the
  target's lightcurve.  $A^\star$ is smaller than the lightcurve
amplitude in all cases and agrees with the actual amplitude within
3$\sigma$ in 3 cases. This behavior is expected since we usually cover
only part of the target rotation with our UKIRT observations. Based on
this assessment, we believe that $A^\star$ and $\tau^\star$ are useful
parameters that provide some constraints on the target's lightcurve
behavior.

\begin{figure}
 \centering
 \includegraphics[width=\linewidth]{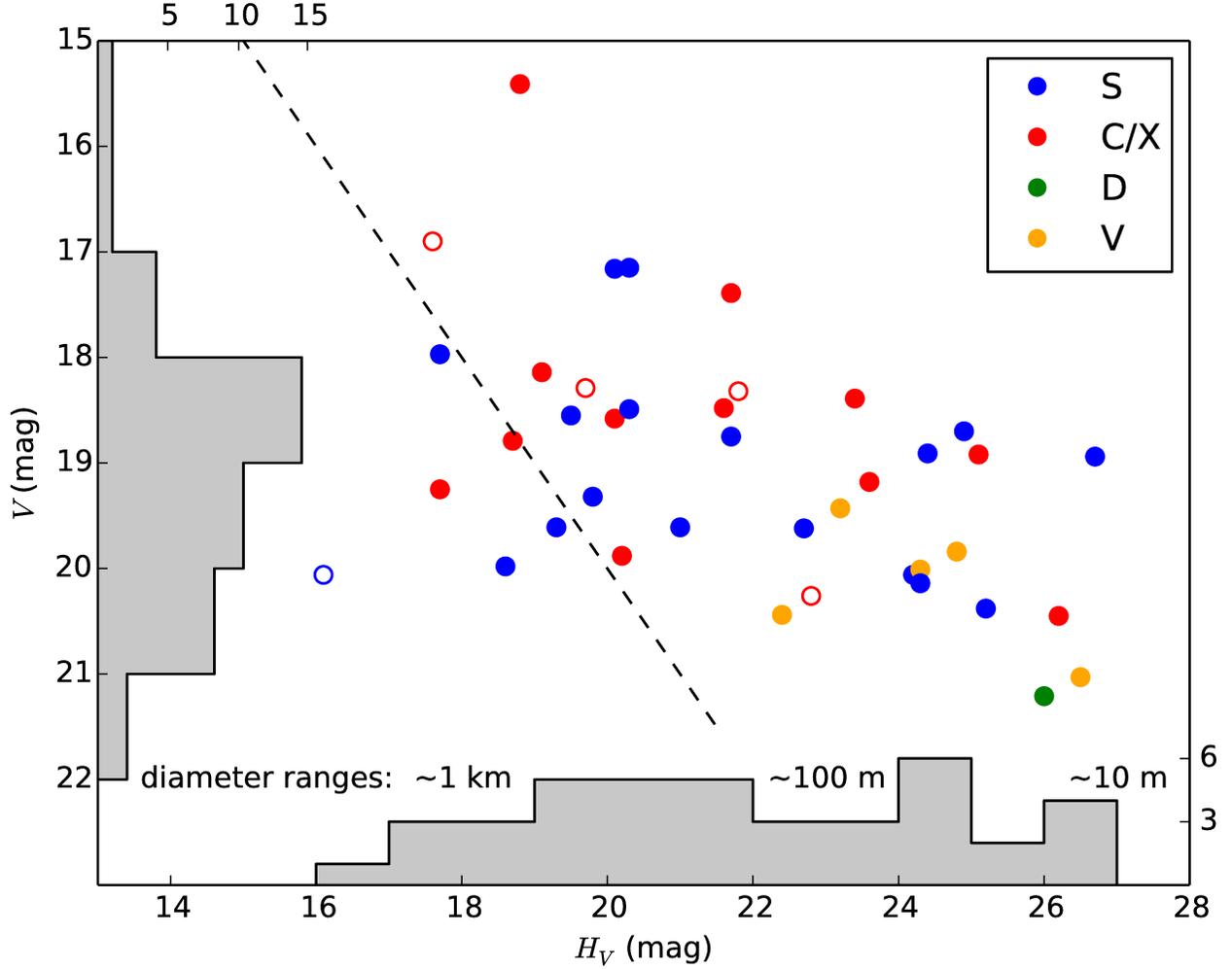}
 \caption{Distribution of sample targets as a function of their
   absolute magnitude $H_V$ and their apparent magnitude $V$ at the
   time of observation; $V$ magnitudes are predicted magnitudes from
   the JPL Horizons system. We only show those targets with reliable
   taxonomic classifications, i.e., with taxonomy probability
   ${\geq}50$\% and root-mean-square color uncertainties
   ${\leq}0.3$~mag. Filled circles represent targets that were
   observed within 4 weeks of discovery; open circles are substitute
   targets that were observed due to the unavailability of rapid
   response targets. The bottom of the plot shows a histogram of the
   $H_V$ magnitude distribution of our targets, the majority of which
   have diameters smaller than 1~km; the vertical axis shows a
   histogram of the $V$ magnitudes. The dashed line represents $V=H_V$;
   most of our targets have been observed when $V<H_V$. We are able to
   derive reliable taxonomic classifications for targets with $V <
   21$~mag.}
 \label{fig:magnitudes}
\end{figure}

We qualitatively investigate the performance of our rapid-response
observing approach. Figure \ref{fig:magnitudes} shows the distribution
of our sample targets as a function of their absolute magnitude $H_V$
and their apparent magnitude $V$ at the time of observation. We only
show those targets for which we could obtain reliable taxonomic
classifications and for which the observations do not suffer from
irregularities (notes in Table \ref{tbl:results}). 90\% of our
reliable sample targets have $H_V>18$~mag, corresponding to asteroid
diameters ${\lesssim}1$~km; 38\% are smaller than ${\sim}100$~m
($H_V\geq23$~mag). Most of our rapid-response targets (excluding
substitute targets) have been observed
at $V<H_V$, which indicates an efficient way to study small
asteroids. Our smallest target, 2014~HW, has $H_V$=28.4~mag, but we
observed it at V=18.9~mag ($H_V-V\sim9.5$) as a result of our
rapid-response observing approach. 2014~HW is not shown on Figure
\ref{fig:magnitudes}, as it is affected by lightcurve
inconsistencies. This target most likely has a diameter of only a few
meters \citep[based on albedo estimates for S-type asteroids
from][]{Thomas2011}, and our comparison to MANOS data (above) confirms 
that we can derive a reliable classification for such small
targets. Figure \ref{fig:magnitudes} shows that we are able to obtain
reliable data for NEOs with $V\leq21$~mag, which is fainter than
targets accessible to spectroscopic observations, except for those
done with the largest telescopes.

\begin{figure}
 \centering
 \includegraphics[width=\linewidth]{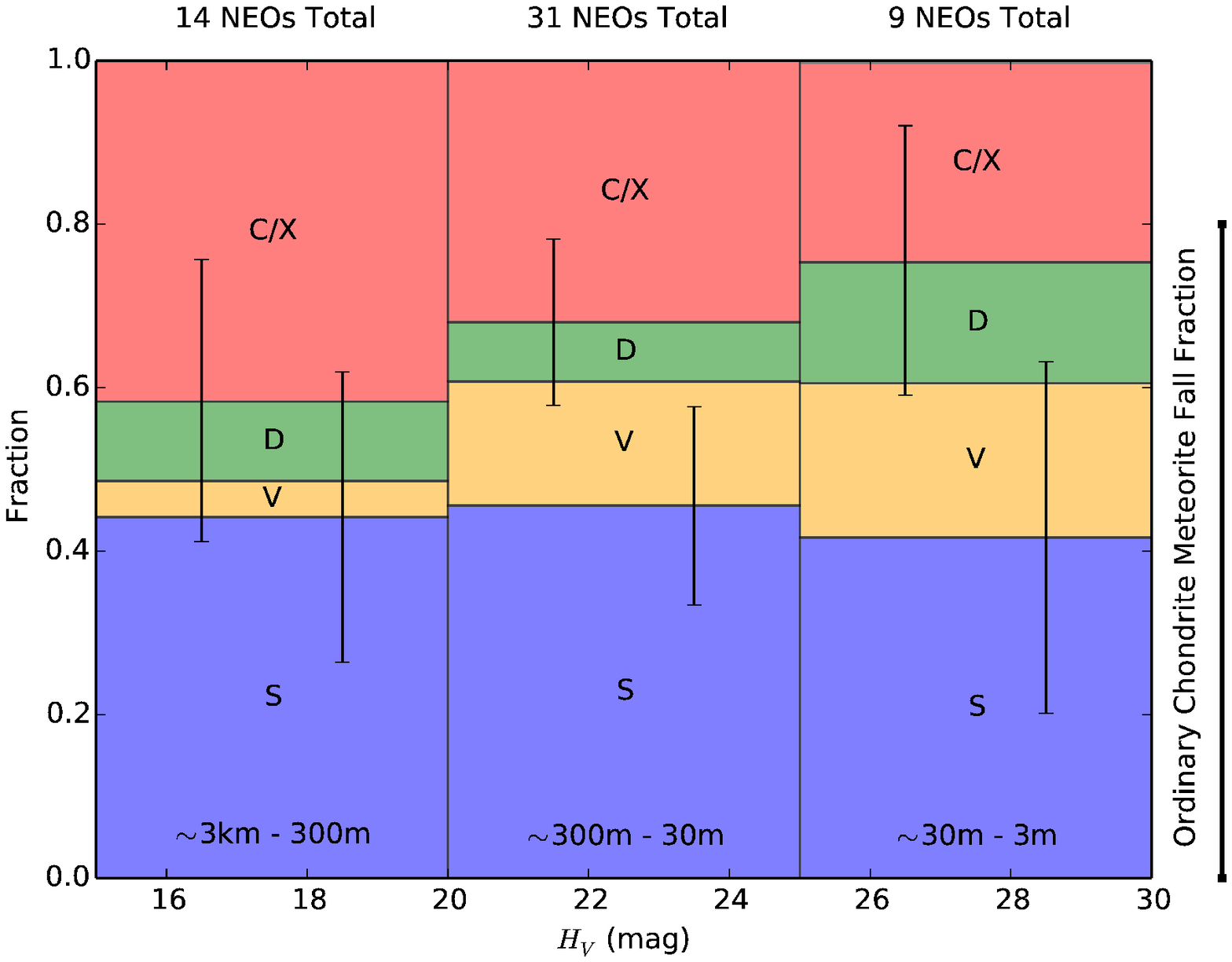}
 \caption{Compositional distribution of our target sample as a
   function of $H_V$, which serves as a proxy for target size (size
   scale assumes an average albedo of 0.15). With decreasing size, the
   fraction of C/X-type asteroids decreases, whereas the fraction of
   D-type and V-type asteroids seems to increase; based on Poisson
   statistics (error bars indicate 1$\sigma$ levels), both trends seem
   to be insignificant.  The fraction of S-type asteroids is
   consistently lower than the expected ${\sim}80$\% level derived
   from ordinary chondrite meteorite falls statistics
   \citep{Harvey1989}; S-type statistics agree with the 80\% level at
   2--3$\sigma$ for all size bins.}
 \label{fig:composition}
\end{figure}

Based on our results in Table \ref{tbl:results}, we examine the
compositional distribution of our sample as a function of absolute
magnitude, which is used here as a proxy for the targets' sizes. For
this analysis, we only take into account sample targets with
root-mean-square color uncertainties ${\leq}0.3$ that have not been
marked in Table \ref{tbl:results} for potential irregularities and
cumulate the individual taxonomic classification
probabilities. Taxonomic classification probabilities for targets with
multiple observations have been averaged.  We compare the fractions of
individual taxonomic types in Figure \ref{fig:composition} in three
different $H_V$-ranges: $15<H_V\leq20$~mag (14 objects),
$20<H_V\leq25$~mag (31 objects), and $25<H_V\leq30$ (9 objects). With
decreasing size (increasing $H_V$), we find a decreasing fraction of
C/X-type NEOs and a relative increase in the fraction of D-type and
V-type NEOs. Dealing with small sample sizes, we apply Poisson
statistics to investigate the significance of these trends and for the
sake of simplicity only consider the S and C/X complexes, which have
the largest sample sizes. Figure \ref{fig:composition} shows that the
1$\sigma$ uncertainties for S and C/X complex objects overlap in each
size bin, which renders the decreasing trend of C/X complex objects
with size insignificant.  We also find the S-complex fraction to be
independent of size at a level of 45\%, which is in stark contrast to
the finding that most meteorite falls are ordinary chondrites
\citep[${\sim}80$\%,][]{Harvey1989}; ordinary chondrites are thought
to originate from S-complex asteroids
\citep{Nakamura2011}. \cite{Hinkle2014} independently find a similar
discrepancy in MANOS data, suggesting a decreasing fraction of S-type
asteroids with decreasing size. Taking the Poisson uncertainties into
account, our S-complex fraction agrees at a 2--3$\sigma$ level with
the fraction of ordinary chondrites for all size bins. Additional data
will be necessary to improve statistics and investigate if this
discrepancy is real. We note that our results do not account for bias
inherent to our target sample. Being optically discovered, our sample
targets are more likely to have medium to high surface albedos than
low albedos, which favors S-complex over C/V-complex asteroids
\citep[e.g.,][]{Thomas2011}. However, this effect would lead to an
overestimation of the fraction of S-complex asteroids, suggesting that
their real fraction is even lower. We postpone a detailed
investigation of the compositional distribution of NEOs and a proper
de-biasing of the distribution as a function of size to future work,
which will be based on a larger sample.

We investigate the possibility that our rapid-response target selection
introduces additional bias into our target sample, favoring specific
taxonomic types. We test this hypothesis by comparing the
distributions in semimajor axis, eccentricity, and inclination for our
target sample and the sample of known NEOs (as reported by the
  Minor Planet Center as of 12 December 2015) using a two-sample
Kolmogorov-Smirnov test. For each of the three parameters we find a
$p$-value $> 0.45$, indicating that we cannot reject the hypothesis
that the distributions of the two samples are the same. Hence, our
targets' orbital parameter distribution does not significantly differ
from the overall NEO distribution.

\section{Conclusions and Future Outlook}

Using rapid-response observations of NEOs in the near-infrared with
UKIRT we are able to estimate taxonomic classifications for our sample
targets in a simplified taxonomy scheme.  Out of 110 observations of
104 different NEOs we derived reliable taxonomic classifications for
46 observations of 43 different NEOs; 18 observations had to be
rejected because the target was too faint or the background too
crowded, 25 more observations are subject to irregularities, 18
observations suffer from a low signal-to-noise ratio (color
uncertainties ${>}0.3$~mag), and 7 observations lead to ambiguous
taxonomic classifications (probabilities for all taxonomic
classifications ${<}50$\%). We expect a significantly higher
efficiency in future observations after adjusting integration times.
We find a good agreement between our taxonomic classifications and
those derived from spectroscopic observations. We are able to reliably
classify NEOs with $V\leq21$~mag, which allows us to characterize
asteroids down to a few meters in diameter, using our rapid-response
approach. Our currently available data sample suggests that the
fraction of S-complex asteroids in the NEO population is lower than
the fraction of ordinary chondrites in meteorite fall statistics.

We will continue our observations with UKIRT and adapt
our observing strategy based on the results of this work. In order to
minimize the effect of galactic extinction on our $Z$ band calibration
(see Section \ref{lbl:zband_calibration}), we refrain from observing
targets at galactic latitudes $|b|<15$\degr. Furthermore, we make $H$
band observations an integral part of our observations, as they
provide additional information that improve the quality of our
taxonomic classification, especially in cases in which a proper $Z$
band calibration is not possible from 2MASS data due to high levels of
galactic extinction. We also adjust total integration times and filter
sequences to provide the necessary photometric accuracy to minimize
the classification uncertainties and improve the sensitivity of our
survey.

\acknowledgments

M.\ Mommert would like to thank the UKIRT and CASU support staff,
including, but not limited to, Watson Varricatt, Peter Milne, Tom
Kerr, and Mike Irwin for their support in the observation planning and
data processing. We thank an anonymous referee for useful suggestions.
This material is based upon work supported by the
National Aeronautics and Space Administration under Grant
No. NNX15AE90G issued through the SSO Near Earth Object Observations
Program.
E.\ Petersen would like to thank the REU program at Northern Arizona
University supported through NSF grant AST-1461200.
The United Kingdom Infrared Telescope (UKIRT) is supported by NASA and
operated under an agreement among the University of Hawaii, the
University of Arizona, and Lockheed Martin Advanced Technology Center;
operations are enabled through the cooperation of the Joint Astronomy
Centre of the Science and Technology Facilities Council of the
U.K. When part of the data reported here were acquired, UKIRT was
operated by the Joint Astronomy Centre on behalf of the Science and
Technology Facilities Council of the U.K. Some of the data reported
here were obtained as part of the UKIRT Service Programme.
Part of the data utilized in this publication were obtained
and made available by the MIT-UH-IRTF Joint Campaign for NEO
Reconnaissance. The IRTF is operated by the University of Hawaii under
Cooperative Agreement no.\ NCC 5-538 with the National Aeronautics and
Space Administration, Office of Space Science, Planetary Astronomy
Program. The MIT component of this work is supported by NASA grant
09-NEOO009-0001, and by the National Science Foundation under Grants
Nos.\ 0506716 and 0907766.  Any opinions, findings, and conclusions or
recommendations expressed in this material are those of the authors
and do not necessarily reflect the views of NASA or the National
Science Foundation.
Taxonomic type results presented in this work were determined in part
using a Bus-DeMeo Taxonomy Classification Web tool by Stephen
M. Slivan, developed at MIT with the support of National Science
Foundation Grant 0506716 and NASA Grant NAG5-12355.



{\it Facilities:} \facility{UKIRT}




\clearpage




\clearpage

\end{document}